\documentstyle[12pt]{article}
\begin{document}
\begin{flushright}
KANAZAWA-00-02 \\ 
March, 2000
\end{flushright}
\vspace*{2cm}
\begin{center}
{\LARGE\bf Radiative symmetry breaking and Higgs mass bound
in the NMSSM}\\
\vspace{1 cm}
{\Large  Y. Daikoku}
\footnote[1]{e-mail: daikoku@hep.s.kanazawa-u.ac.jp}
 ~{\Large and}~ {\Large D. Suematsu}
\footnote[2]{e-mail: suematsu@hep.s.kanazawa-u.ac.jp}
\vspace {1cm}\\
{\it Institute for Theoretical Physics, Faculty of Science,
Kanazawa University,\\
        Kanazawa 920-1192, Japan}
\end{center}
\vspace{2cm}
We study the upper mass bound of the lightest neutral Higgs scalar
in the NMSSM using the RGE analysis. 
We require the successful occurence of the electroweak radiative 
symmetry breaking to restrict the parameter space. 
As a result the upper mass bound $m_{h^0}$ is largely restricted
compared with the one estimated without imposing this condition.
We point out some features of $m_{h^0}$ 
related to the initial value of $h_t$
and discuss why the models with more extra matters ${\bf 5}+\bar{\bf
5}$ of SU(5) could bring the larger maximum value of $m_{h^0}$. 

\newpage
\noindent

The existence of a rather light CP-even neutral Higgs scalar is a common 
feature of the supersymmetric extension of the standard model
\cite{higgs1,higgs2}. It is
an important evidence of this kind of models and it is very useful to
know its possible upper bound for the judgement of the consistency of
the models.

The next to the minimal supersymmetric standard model (NMSSM)
is the simplest extension of the minimal supersymmetric standard
model (MSSM) \cite{nmssm}. In this model a singlet chiral 
supermultiplet $S$ is 
introduced and a $\mu$ term in the MSSM is replaced by a Yukawa coupling
$\lambda SH_1H_2$ with the usual Higgs doublet chiral superfields $H_1$ 
and $H_2$. The superpotential of the Higgs sector in this model 
is expressed as
\begin{equation}
W_{\rm NMSSM}=\lambda SH_1H_2 + {1 \over 3}\kappa S^3 +\cdots.
\end{equation}
If the scalar component $\tilde S$ of $S$ gets a vacuum expectation
value, the $\mu$ scale appears as $\lambda \langle \tilde S\rangle$.
In this model the $\mu$ problem in the MSSM is potentially solvable 
when the tree
level $\mu$ term is forbidden due to a suitable symmetry \cite{mu}.

An interesting feature of this model is the fact that the mass bound
of the lightest neutral Higgs scalar can be estimated with no dependence
on the soft supersymmetry breaking parameters at the tree level \cite{bound},
\begin{equation}
m_{h^0}^{(0)2} \le m_Z^2\left[\cos^22\beta +{2\lambda^2\over
g_1^2+g_2^2}\sin^22\beta\right], 
\end{equation}
where $\tan\beta=\langle H_2\rangle/\langle H_1\rangle$.
This bound is mainly controled by the value of $\tan\beta$ and the
bound of $\lambda$. Its dependence on the soft supersymmetry breaking
parameters appears through the loop correction to the effective
potential as a result of the large top Yukawa coupling $h_t$ \cite{higgs2}.
All of these effects are related through the renormalization
group equations (RGEs).
Many works have been done on this aspect \cite{nmssm2,nmssm3}.
In Ref.\cite{nmssm3} it has been suggested that the additional 
extra matters such as ${\bf 5}+\bar {\bf 5}$ of SU(5) can heavily 
affect the mass bound by changing the running of gauge couplings, $h_t$ 
and $\lambda$ but
not affecting the scale of the gauge coupling unification, which is a great
success of the MSSM. 

In the NMSSM the radiative generation of nonzero $\langle \tilde
S\rangle$ or the $\mu$ scale is a very important aspect. 
From this point of view 
it seems to be rather crucial to study how the requirement of 
the successful occurence of the radiative symmetry breaking \cite{rad} 
affects on the upper mass bound of the lightest neutral Higgs scalar. 
In this note we investigate this problem by finding the radiatively induced 
minimum of the effective potential parameterized by 
$(\tan\beta, \langle \tilde S\rangle)$ in the models with 
extra matters $n({\bf 5}+\bar {\bf 5})$ of SU(5).
 
The one-loop effective potential due to the large top Yukawa coupling
\cite{effective,effective1} is,
\begin{equation}
V_1={1\over 64\pi^2}\left[-12m_t^4\left(\ln{m_t^2\over Q^2}-{3\over 2}\right)
+\sum_{i=1}^26\tilde m_{t_i}^4
\left(\ln{\tilde m_{t_i}^2\over Q^2}-{3\over 2}\right)\right],
\end{equation} 
where $Q$ is a renormalization point and 
$\tilde m_{t_i}^2$ is the eigenvalue of the stop mass matrix
\begin{equation}
\left(\begin{array}{cc}
\tilde m^2_{Q} +m^2_t & m_t(-A_t + \lambda \langle \tilde S\rangle\cot\beta)\\
 m_t(-A_t + \lambda \langle \tilde S\rangle\cot\beta) & 
\tilde m^2_{\bar T} +m^2_t \\
\end{array}\right).
\end{equation}
The correction to Eq. (1) due to this one-loop effective potential
is expressed by using these mass eigenvalues as 
\begin{equation}
\Delta m_{h^0}^2={1\over 2}\left({\partial^2 V_1 \over\partial
v_1^2}-{1\over v_1}{\partial V_1\over \partial v_1}\right)\cos^2\beta
+{1\over 2}{\partial^2 V_1 \over\partial v_1\partial v_2}\sin 2\beta
+{1\over 2}\left({\partial^2 V_1 \over\partial v_2^2}
-{1\over v_2}{\partial V_1\over \partial v_2}\right)\sin^2\beta.
\end{equation}
Here we used the potential minimum condition to eliminate the soft
scalar masses of Higgs fields and took a field basis which is used to
derive Eq. (2).
The mass matrix in Eq. (4) depends on $(\tan\beta, \langle\tilde
S\rangle)$ other than the Yukawa couplings and the soft 
SUSY breaking parameters.
  
In order to fix this matrix we need to determine these values as the
ones at the
potential minimum. It is a nontrivial problem whether such values of 
$(\tan\beta, \langle\tilde S\rangle)$ can
be radiatively realized starting from certain sets of the Yukawa
couplings and the soft SUSY breaking parameters.
Our task is to estimate $m_{h^0}^2(\equiv m_{h^0}^{(0)2}+\Delta
m^2_{h^0}) $ numerically for the parameter sets which can 
radiatively realize the phenomenologically acceptable potential
minimum.
We determine such parameter sets so as to satisfy the following
conditions:\\
(i)~starting from the suitable initial values of parameters, 
the radiative symmetry breaking occurs successfully and the following 
phenomenologically required condition is satisfied at the potential minimum,\\
\begin{equation}
\langle H_1\rangle^2+\langle H_2\rangle^2=(174~{\rm GeV})^2, 
\qquad m_t=174~{\rm GeV},
\end{equation}
(ii)~$m_{h^0}^{2}$ which corresponds to the one of diagonal elements
of the $3\times 3$ neutral Higgs mass matrix should be smaller than other
two diagonal components \cite{bound},\\
(iii)~the experimental mass bounds on the charged Higgs bosons $m_{H^\pm}$, 
charginos $m_{\chi^\pm}$, stops $\tilde m_{t_i}$ and gluinos $M_3$ are 
satisfied. 
These masses, except for $M_3$, are dependent on $\lambda$ 
and $\langle \tilde S\rangle$.
We require the following values for them:
\begin{equation}
m_{H^\pm} > 65~{\rm GeV},\quad m_{\chi^\pm} >72~{\rm GeV} \quad
\tilde m_{t_2} > 67~{\rm GeV},\quad
 M_3 > 173 ~{\rm GeV},
\end{equation}
(iv)~the vacuum should be a color conserving one \cite{colorb}.
  
In this study we solve a set of RGEs which are composed of 
two-loop ones for dimensionless couplings and one-loop ones for 
dimensional SUSY breaking parameters, for simplicity.
As the initial conditions for the SUSY breaking parameters we take
\begin{equation}
\tilde m_{\phi_i}^2=(\gamma_i\tilde m)^2, \qquad M_a=M, 
\qquad A_t=A_\kappa =A_\lambda =A,
\end{equation}
where $\tilde m$ is the universal soft scalar mass. 
We introduce the nonuniversality represented by $\gamma_i$ only among
soft scalar masses of $H_1, H_2$ and $S$ to make it easy to find the
radiative symmetry breaking solutions. 
This will be taken as $0.8 \le \gamma_i \le 1.2$. 
These initial conditions are assumed to be applied at the scale $M_X$ 
where the 
coupling unification of SU(2)$_L$ and U(1)$_Y$ occurs. We donot require
the regolous coupling unification of SU(3)$_C$ but only impose the
realization of the low energy experimental value \cite{nmssm3}.
The initial values of the parameters are surveyed through 
the following region,
\begin{eqnarray}
&&0\le h_t\le 1.2~~(0.1),\qquad -2.0\le \kappa\le 0~~(0.2), 
\qquad 0\le \lambda \le 3.0~~(0.2), \nonumber\\
&&0\le M/M_S \le 0.8~~(0.3), \qquad
0\le \tilde m/M_S,~ |A|/M_S \le 3.0~~(0.5),
\end{eqnarray}
where in the parentheses we give the interval which we use in the
survey of these parameter regions.\footnote{
Since the sign of $\kappa$ and $A$ affects the scalar potential, we
need to investigate both sign of them.
However, a negative $\kappa$ seems to cover almost solutions for the
positive $\kappa$. Here we give the only result in the case of the
negative $\kappa$.}
We also assume that the RGEs of the model are changed from the
supersymmetric ones to the nonsupersymmetric ones at a supersymmetry 
breaking scale $M_S$, for which we take $M_S=1$~TeV as a typical numerical 
value \cite{nmssm2,nmssm3}.

To estimate the one-loop effect Eq. (5), it is necessary 
to know the values of $(\tan\beta,\langle\tilde S\rangle)$ 
at the potential minimum.
If we impose the radiative symmetry breakling condition, such a
potential minimum has to be realized as a result of RGEs solution.
In order to see the effect of the radiative 
symmetry breaking condition on the upper mass bound $m_{h^0}$ 
we calculate it under two situations.
We impose the full condition of (i) in a case (I).
On the other hand, in a case (II) we require only Eq. (6) but donot
require that it is realized at the potential minimum\footnote{
In other words, this case corresponds to the situation that
Eq. (6) may be satisfied at the minimum which is obtained from the 
unnatural initial soft scalar masses.}.
We take the number of extra matters as $n=3$. 
The perturbative unification of the gauge coupling requires $n\le 4$. 
Later we will also discuss other cases.

At first, we show how the radiative symmetry breaking condition restricts the
parameters strongly relevant to $m_{h^0}$.  In Fig. 1 we give the plots of the
solutions in $(\tan\beta,\langle \tilde S\rangle)$ and
$(\tan\beta,\lambda)$ planes. Since we assume that only the top Yukawa
coupling is large, the consistent $\tan\beta$ cannot be so
large\footnote{To estimate $\tan\beta$ we take account of the 
translation of the pole mass to the running mass \cite{run}.}. 
We take it as $\tan\beta \le 15.0$.
In the case (II), $\langle \tilde S\rangle$ can take the value in such a
wide range as $70~{\rm GeV}~{^<_\sim}~\langle \tilde S\rangle~
{^<_\sim}~45~{\rm TeV}$, which is not plotted in Fig. 1(a).
These show that the value of $\lambda(m_t)$ and 
$\langle \tilde S\rangle$ 
are heavily restricted in the case (I) compared with the case (II).
We should remind that $\lambda(m_t)$ and $\langle \tilde S\rangle$ 
affect $m_{h^0}$ through Eqs. (2) and (4).
As seen from Fig. 1, the $\mu(\equiv \lambda\langle \tilde S\rangle)$
scale can take a value in a rather wide range to cause the successful
radiative symmetry breaking.

\input epsf 
\begin{figure}[tb]
\begin{center}
\epsfxsize=8cm
\leavevmode
\epsfbox{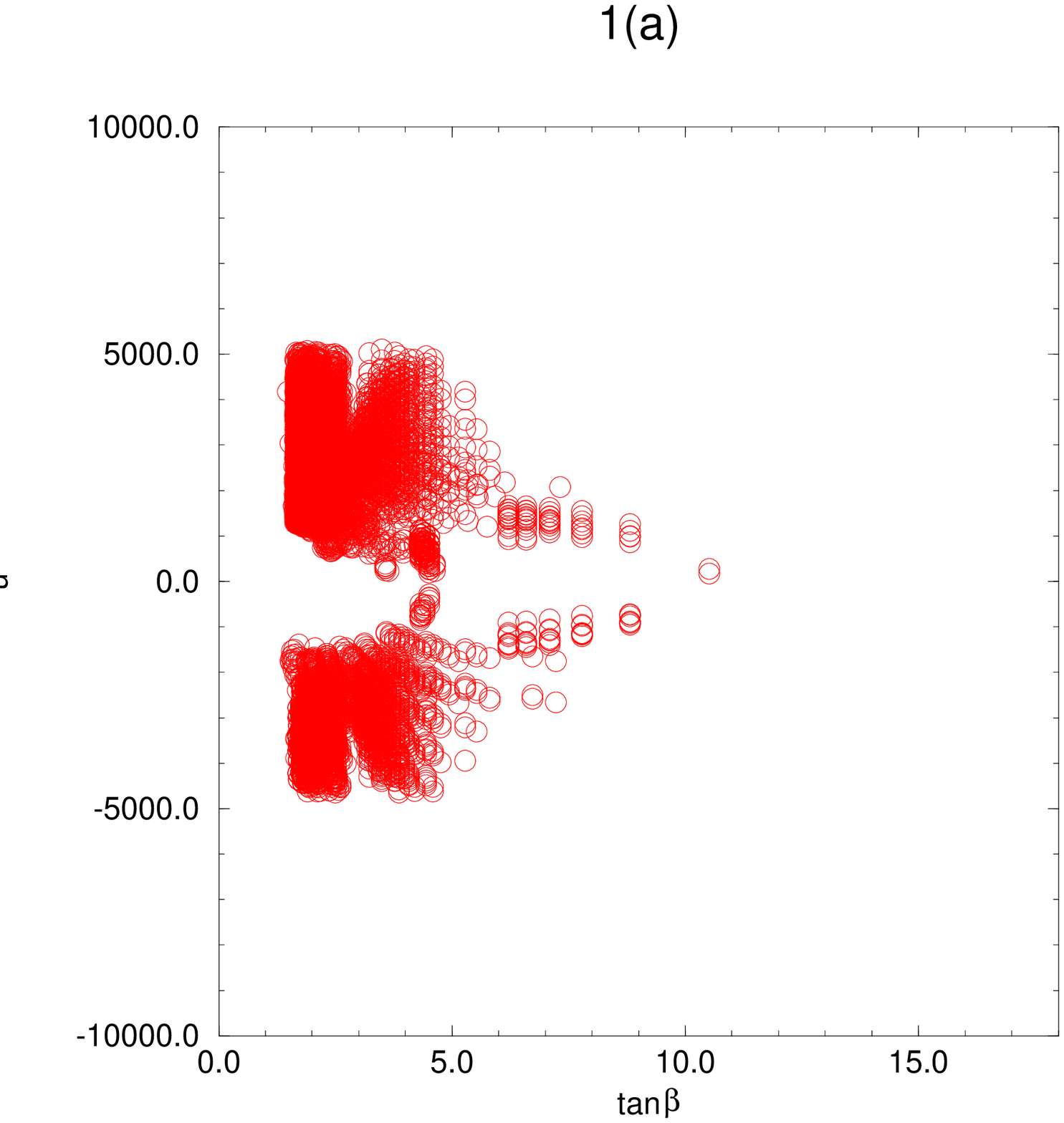}
\hspace*{-0.3cm}
\epsfxsize=8cm
\leavevmode
\epsfbox{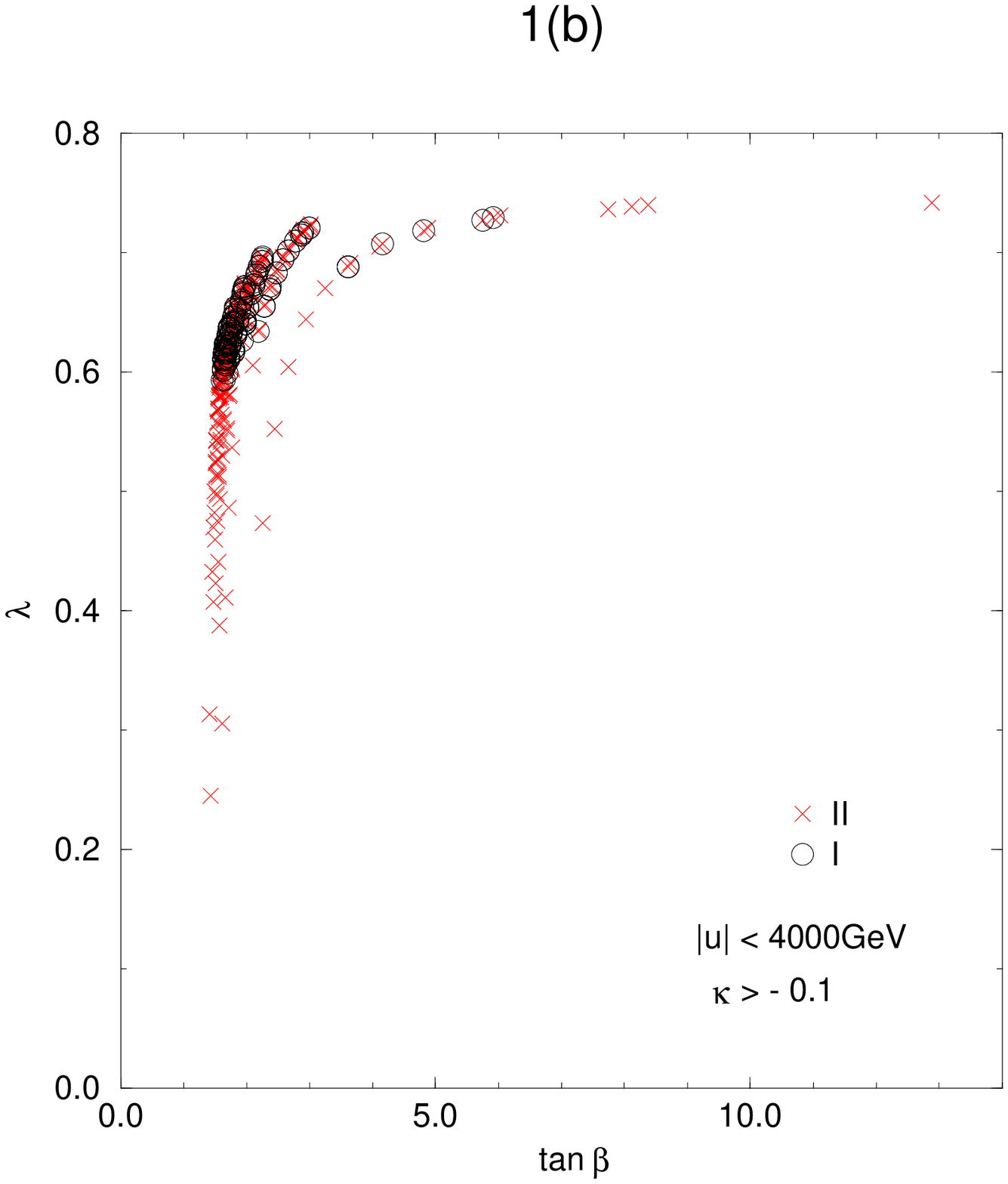}
\end{center}
\vspace*{-1cm}
{\footnotesize {\bf Fig. 1}~~\  Scatter plots of the solutions for the case
(I) and the case (II) in $(\tan\beta, u(\equiv\langle \tilde S\rangle)$ and
$(\tan\beta,\lambda)$ planes at the $m_t$ scale. In 1(a) 
only the solutions in (I) are drawn by circles.}
\end{figure}

In Fig. 2 we show the allowed region of $m_{h^0}$ in both cases 
for the corresponding 
parameter sets to the ones of Fig. 1.  
This shows that the imposition of the consistent occurence of 
the radiative symmetry breaking can strongly affect the estimation of
$m_{h^0}$. The boundary value of $m_{h^0}$ can be changed by a few
percent to ten percent.
We can also find some parameter dependences of $m_{h^0}$ in these figures. 
In Fig. 2(b) we restrict the initial soft scalar mass as 
$\tilde m=1$~TeV.  
The larger values of soft scalar mass $\tilde m$ and 
$\langle \tilde S\rangle$ tend to realize the larger value of $m_{h^0}$. 
The value of $\langle \tilde S\rangle$ determines the off-diagonal 
component of the stop mass matrix (4). This shows that the larger stop 
mixing tends to make $m_{h^0}$ larger.  

\begin{figure}[tb]
\begin{center}
\epsfxsize=8cm
\leavevmode
\epsfbox{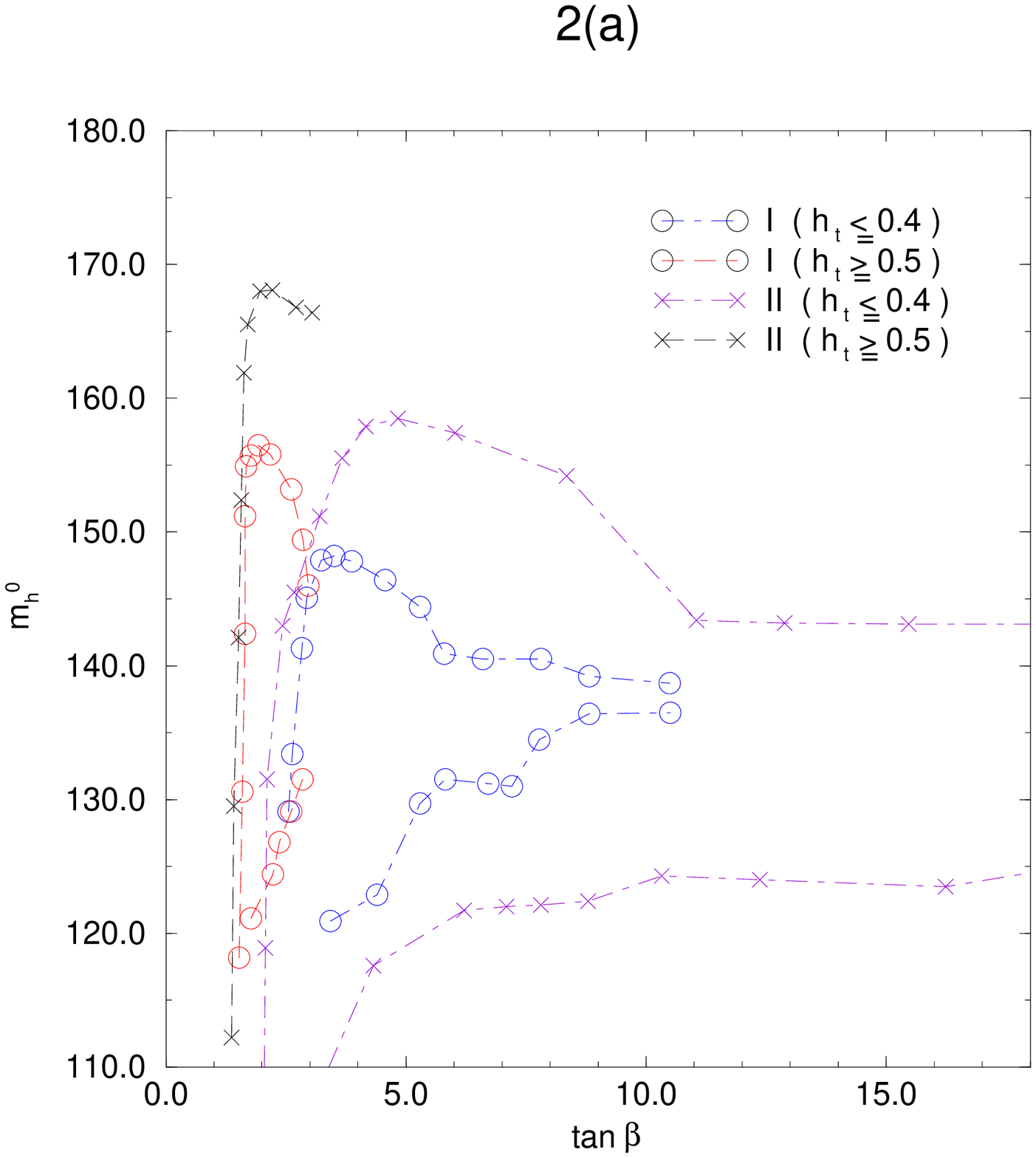}
\hspace*{-0.3cm}
\epsfxsize=8cm
\leavevmode
\epsfbox{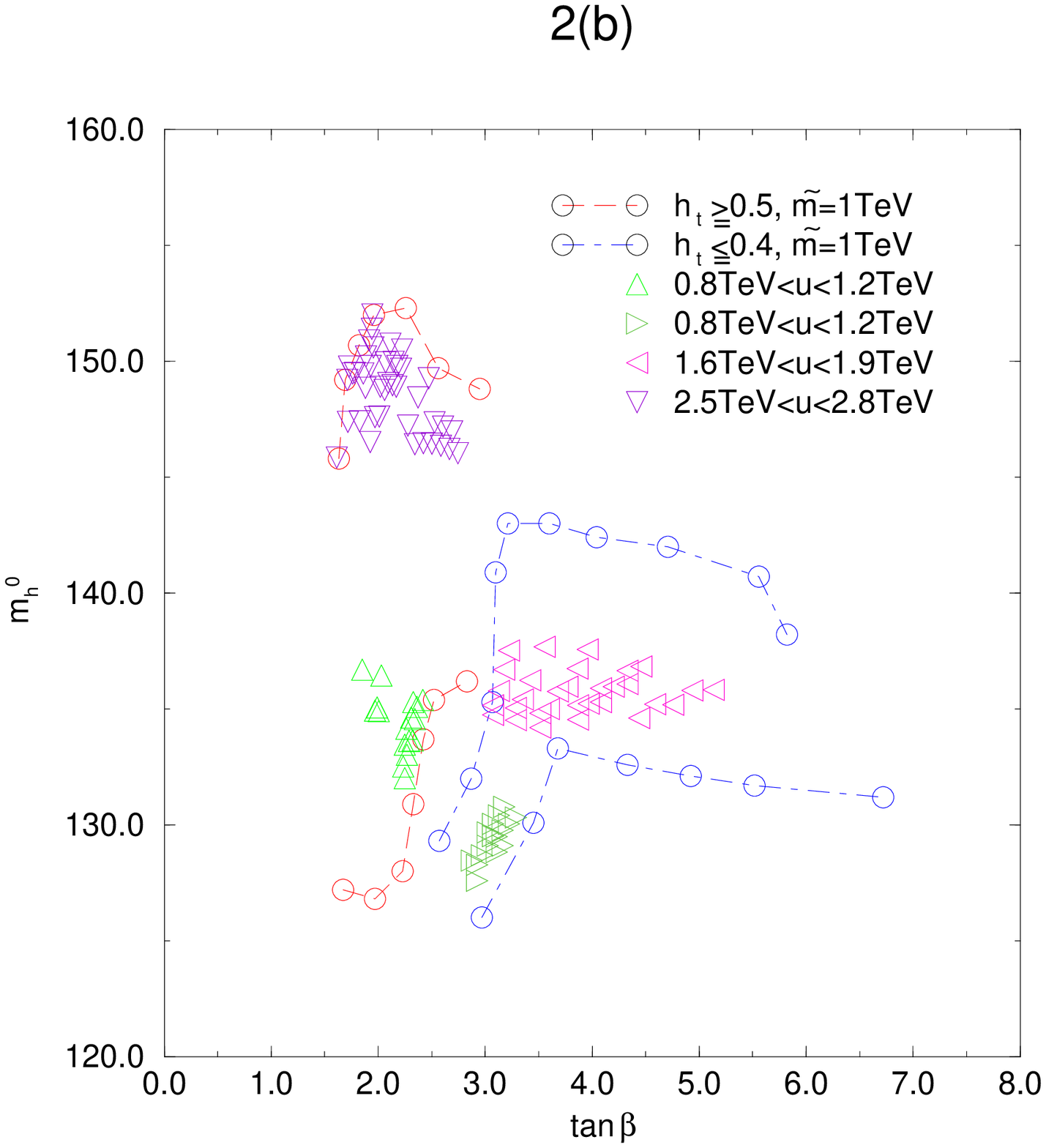}
\vspace*{-1cm}
\end{center}
{\footnotesize {\bf Fig.2}~~\  Allowed regions of $m_{h^0}$.
The boundary values are shown by the dashed and dash-dotted lines for
$h_t \ge 0.5$ and $h_t\le 0.4$, respectively.
The displayed values of the top Yukawa coupling $h_t$ and the soft scalar mass 
$\tilde m$ are the ones at the unification scale $M_X$.}
\end{figure}

It is useful to comment on the existence of two branches of the
solutions which show a very different behavior in Figs. 1 and 2. 
This is a common feature in both case (I) and (II) and also in the
models with a different $n$. For simplicity, we take the case (II) as 
an example to discuss this feature here.
We refer the one of the smaller $\tan\beta$ as a branch B1 and the one
of the larger $\tan\beta$ as a branch B2. 
They are divided by an initial value
of $h_t$ as shown in Table 1. 
The interesting feature of two branches can be clearly seen in
Fig. 1(b) and Table 1.
The branch B2 corresponds to the smoothly extending
solutions of $\lambda(m_t)$ to the large $\tan\beta$ region and the branch B1 
comes from the ones which are confied in the small $\tan\beta$.
The branch B2 has a similar lower bound
of $\tan\beta$ and the similar maximum value of $m_{h^0}\sim 143$GeV
at $\tan\beta~{^>_\sim}~9$ for the different $n$ values.
If we note that at the larger $\tan\beta$ region the second term of
Eq. (2) can be neglected and Eq. (2) reduces to the one of the MSSM,
this behavior can be understood. 
On the other hand, the branch B1 strongly depends on $n$. There are solutions
with a little bit larger maximum value of $\lambda$ at 
the smaller $\tan\beta$ according to the increase of $n$.   
This results in the larger maximum value of $m_{h^0}$ for the larger
$n$. The reason can be mainly found in the $\beta$ dependence of Eq. (2). 
Decreasing $n$, the maximum value of $\tan\beta$ of the branch B1 
increases, where the maximum value of $\lambda(m_t)$ is realized. 
The branch B1 disappears finally at $n=0$ since it is difficult
to realize $0.95~{^<_\sim}~h_t(m_t)~{^<_\sim}~1.35$
corresponding to $1~{^<_\sim}~\tan\beta~{^<_\sim}~15$ starting from 
the initial value of $h_t$ used here. 
We should remind that the value of $h_t(m_t)$ is strictly restricted 
by  Eq. (6). 
In the case (I) we also find the similar qualitative feature discussed 
here in the case (II). 

Finally we want to present one comment.
There is a rather big difference of the density of the
radiative symmetry breaking solutions in the models with the different 
$n$. For example, in the present parameter setting the number of
solutions in $n=4$ is very smaller than the ones in $n=3$.
The finer tuning of marameters seems to be necessary to find the
solution in the $n=4$ case compared with the $n=3$ case.

\begin{figure}[tb]
\begin{center}
\begin{tabular}{|c|c|c|c|c|c|c|c|c|}\hline
  & \multicolumn{4}{|c|}{branch B1} & \multicolumn{4}{|c|}{branch B2}
  \\ \hline
n & $h_t(M_X)$ & $\tan\beta$ & $\lambda^{\rm max}(m_t)$ &
$m_{h^0}^{\rm max}$&$h_t(M_X)$ & $\tan\beta$ & $\lambda^{\rm max}(m_t)$ 
&$m_{h^0}^{\rm max}$\\ \hline\hline
0 & --   &  --   &  --  & --   & $\sim$1.2 & 2.1$\sim$  & 0.69& 149 \\
1 &0.9$\sim$1.2& 1.8$\sim$6.3 & 0.70 & 161& 0.6$\sim$0.8&2.3$\sim$   
& 0.70 & 145\\
2 &0.8$\sim$1.2& 1.6$\sim$4.5 & 0.71 & 165 &0.5$\sim$0.7& 2.2$\sim$ 
& 0.72   &149\\
3 &0.5$\sim$1.2& 1.4$\sim$3.0 & 0.72 & 168 &0.3$\sim$0.4& 2.1$\sim$  
& 0.75  &158\\
4 &0.2$\sim$0.1& 1.2$\sim$1.8 & 0.69 & 170 & $\sim$0.1    & 2.3$\sim$  
& 0.73  & 149\\ \hline
\end{tabular}
\vspace*{3mm}

{\footnotesize {\bf Table 1}~~\ Extra matter effects and the branch
structure in the case (II). }
\end{center}
\end{figure}

In summary we studied the upper bound $m_{h^0}$ of the lightest neutral Higgs
scalar mass in the NMSSM using the RGE analysis. 
We required the successful occurence of the electroweak radiative 
symmetry breaking starting from the suitable initial values of
parameters. This condition substantially 
constrains the allowable parameter
space and as a result the mass bound $m_{h^0}$ is heavily restricted
compared with the one obtained without imposing this condition.
We discussed the typical feature related to the initial value of $h_t$
and also why the larger $n$ models could bring the larger maximum value of 
$m_{h^0}$. 

\vspace*{0.3cm}
This work has been partly supported by the
a Grant-in-Aid for Scientific Research from the Ministry of Education, 
Science and Culture(\#11640267 and \#11127206).

\end{document}